\documentstyle[aps,prl,psfig,epsf,epsfig]{revtex}
\newcommand{\istyle}{\mbox{$i$}}

\begin{document}
\twocolumn
\draft
\title{Josephson effects in dilute Bose-Einstein
condensates}
\author{S. Giovanazzi\cite{sg}, A. Smerzi and S. Fantoni}
\address{
Istituto Nazionale di Fisica della Materia and
International School for Advanced Studies,\\
via Beirut 2/4, I-34014, Trieste, Italy,
}
\date{\today}
\maketitle
\begin{abstract}
We propose an experiment that would demonstrate the ``dc'' and
``ac'' Josephson effects 
in two weakly linked Bose-Einstein condensates. We consider  
a time-dependent barrier, moving adiabatically 
across the trapping potential. 
The phase dynamics are governed by a ``driven-pendulum'' 
equation, as in current-driven superconducting Josephson 
junctions. 
At a critical velocity of the barrier (proportional to 
the critical tunneling current), there is a sharp 
transition between the  ``dc'' and ``ac'' regimes. 
The signature is a sudden jump of a large fraction of the 
relative condensate population.
Analytical predictions are compared with 
a full numerical solution of the time dependent Gross-Pitaevskii 
equation, in an experimentally realistic situation.
\end{abstract}

The Josephson effects (JE's) are a paradigm of 
the phase coherence manifestation
in a macroscopic quantum system \cite{anderson84,barone82,barone99}.
Observed early on in superconductors \cite{barone82},
JE's have been demonstrated in two 
weakly linked superfluid $^3$He-B reservoirs 
\cite{pereverzev97}.
Weakly interacting Bose-Einstein condensate (BEC) gases 
\cite{anderson95} provide a further (and different) context 
for JE's.
Indeed, magnetic and optical traps can be tailored and 
biased (by time-dependent external probes) with high 
accuracy \cite{andrews97,hall98,anderson98},  
allowing the investigation of dynamical regimes that might not 
be accessible with other superconducting/superfluid systems.
The macroscopic BEC's coherence has been demonstrated by 
interference experiments \cite{andrews97,hall98}, 
and the first evidence of coherent tunneling in an 
atomic array, related to the ``ac'' JE, has been recently 
reported \cite{anderson98}.

A superconducting Josephson junction (SJJ) is usually 
biased by an external circuit that typically
includes a current drive $I_{ext}$.
The striking signatures of the Josephson effects in SJJ are
contained in the voltage-current characteristic 
($V$-$I_{ext}$),
where usually one can  distinguish between
the superconductive branch or ``dc''-branch 
($V=0$, $I_{ext}\ne 0$),
and the resistive branch or ``ac''-branch
($V \approx R\,I_{ext}$)
(see for example \cite{barone82}).
External circuits and current sources are absent
in two weakly linked Bose condensates
and the Josephson effects have been related, so far, with coherent
density oscillations between condensates in two traps or between
condensates in two different hyperfine levels 
\cite{smerzi97,milburn97,ruo98,villain99,zapata98,williams99}.
This collective dynamical behavior is described by a non-rigid 
pendulum equation \cite{smerzi97}, predicting a new class 
of phenomena not observable with SJJ's.

Now the following question arises:
can two weakly linked condensates exhibit the analog
of the voltage-current characteristic in SJJ?
Although BECs are obviously neutral, the answer is positive.
A dc current-biased SJJ can be simulated by considering a 
tunneling barrier moving with constant velocity 
across the trap.
At a critical velocity of the barrier a sharp transition 
between the ``dc'' and ``ac'' (boson) Josephson regimes occurs.
This transition is associated with a macroscopic jump in 
the population difference, that can be easily monitored 
experimentally by destructive or non-destructive
techniques.

In the following we will briefly introduce the 
phenomenological equations of the 
resistively shunted junction (RSJ) model for the SJJ.
We will describe the corresponding experiment
for two weakly linked BECs and  show 
that the relevant equations are formally equivalent
to the RSJ equations.
Then we compare the analytical results with 
a numerical integration of the Gross-Pitaevskii equation 
in a realistic 3D setup.

In the RSJ model, 
SJJ is described by an equivalent circuit
\cite{barone82} 
in which the current balance equation is
\begin{equation} 
I_c \sin (\theta) + G\,V + C\,\dot{V} = I_{ext} 
\label{eq0}
\end{equation}
where $I_c$ is the upper bound of the Josephson supercurrent $I$
(which is represented, in the ideal case,
by the sinusoidal 
current-phase relation $I = I_c \sin (\theta)$);
$G$ is an effective conductance (offered by the 
quasiparticles and the circuit shunt resistor), 
and $C$ is the junction capacitance.
The voltage difference $V$ across the 
junction is related to the relative phase $\theta$ by
\begin{equation}
\dot{\theta} = 2 e V / \hbar \;. 
\label{eq01}
\end{equation}
In the low conductance limit 
$G \ll\omega_p \,C$  where
$\omega_p = \sqrt{2 e I_c / \hbar C}$
is the Josephson plasma frequency,
combining equations (\ref{eq0}) and (\ref{eq01}) 
leads to the ``driven pendulum'' equation
\begin{equation}
\ddot{\theta} = 
%- {\omega_p} \beta \;\dot{\theta}
- {\omega^2_p} \frac{\partial}{\partial\theta}
U \left( \theta \right) 
\label{eq1}
\end{equation}
where $U$ is the tilted ``washboard'' potential: 
\begin{equation}
U \left( \theta \right) = 1 - \cos(\theta) + \istyle \,\theta 
\label{eq2}
\end{equation}
with $\istyle = I_{ext}/I_c$.
This equation describes the transient behavior 
before the stationary dissipative behavior is reached 
(resistive branch). 
If we start from equilibrium, with $\istyle =0$,
and increase adiabatically the current, no voltage drop
develops until the critical value $\istyle = 1 $ 
is reached (neglecting secondary quantum effects).
At this point $V$ continuously develops until  
a stationary asymptotic dissipative behavior is reached 
in a time scale approximately of order $C/G$.
Similar phenomenology may occur in BECs and we will derive equations
formally identical to Equations (\ref{eq1}) and (\ref{eq2}).

A weak link between two condensates can be created by
focusing a blue-detuned far-off-resonant laser sheet
into the center of the magnetic trap \cite{andrews97}. 
The weak link can be tailored by tuning 
the width and/or the height of the laser sheet. 
Raman transitions between two condensates in different 
hyperfine levels provide a different weak link \cite{hall98}, 
in analogy with the ``internal Josephson effect'' observed 
in $70$s with $^3He-A$ \cite{webb74}.

Here we consider a double well potential in which the laser 
sheet slowly moves across the magnetic trap with velocity $v$
(but our framework can be easily adapted to investigate
the internal Josephson effect).
In the limit of very low $v$, 
the two condensates remain in 
equilibrium, i.e. in their instantaneous ground state,
because of the non-zero tunneling current that can be 
supported by the barrier. 
In fact, an average net current, proportional to the velocity 
of the laser sheet, flows through the barrier, sustained 
by a constant relative phase between the two condensates.
This keeps the chemical potential difference between the two 
subsystems locked to zero, as in the SJJ dc-branch. 
However, the superfluid component of the current 
flowing through the barrier is bounded by a 
critical value $I_c$. 
As a consequence there exists a critical velocity $v_c$, 
above which a non-zero chemical potential difference 
develops across the junction. 
This regime is characterized by a running-phase mode,
and provides the analog of the ac-branch in SJJ's.

The "dc" and "ac" BEC regimes are governed by a 
phase-equation similar to the current-driven pendulum 
equations (\ref{eq1}) and (\ref{eq2}).
Such equations together with the sinusoidal 
current-phase relation $I=I_c\sin(\theta)$ describe 
the phase difference and current dynamics.
The dimensionless current $\istyle$ is related to the barrier velocity by
\begin{equation}
\istyle = v\,/\,v_c
\label{eq4}
\end{equation}
with the critical velocity $v_c$ given by
\begin{equation}
v_c= \frac{\hbar\omega_p^2}F
\label{eq5}
\end{equation}
where $F$ is to a good approximation
represented by double the average 
force exerted by the magnetic trap 
on single atoms in one well.

Equations (\ref{eq1})-(\ref{eq5})
can be derived by a time-dependent variational approximation
and have also been verified, as we discuss below, 
by the full numerical integration  
\cite{giovanazzi} of the Gross-Pitaevskii equation (GPE) 
\cite{pitaevskii61,dalfovo98}.
The GPE describes the collective dynamics of a dilute 
Bose gas at zero temperature:
\begin{equation}
 i \hbar 
\frac \partial {\partial t}\Psi \,\,=
\left[ H_0\left(t\right) +g\,|\Psi |^2\right]
\Psi   
\label{eq6}
\end{equation}
where $ H_0\left(t\right)=
- \frac{\hbar ^2}{2m}{\bf \nabla }^2 
+ V_{ext}\left( {\bf r},t\right)        $
is the non interacting Hamiltonian and 
where $g=4\pi \hbar ^2 a / m$, with $a$ the scattering 
length and $m$ the atomic mass.
The order parameter $\Psi =\Psi\left( {\bf r},t\right) $  
is normalized as 
$\int d{\bf r\,\,}|\Psi \left( {\bf r},t\right) |^2=N$, 
with $N$ the total number of atoms. 
The external potential is given by the magnetic trap and 
the laser barrier $V_{ext}\left( {\bf r},t\right) 
=V_{trap}\left( {\bf r}\right) +V_{laser}\left( z,t\right)$. 
We consider a harmonic, cylindrically symmetric  trap
$V_{trap}\left( {\bf r}\right) =\frac 12 m \omega _r^2
\left( x^2+y^2\right) + \frac 12m\omega _0^2\,z^2$
where $\omega_r$ and $\omega_0$ are the radial and 
longitudinal frequency, respectively. 
The barrier is provided by a Gaussian  shaped laser 
sheet, focused near the center of the trap 
$V_{laser}\left( z\right) = V_0 \exp \left( 
-(z-l_{z}) ^2 / \lambda^2 \right)$ with the coordinate 
$l_{z}(t)$ describing the laser motion 
and $v = d\,l_{z}/dt$ its velocity.

The equations (\ref{eq1}) to (\ref{eq5}) can be  
derived by solving variationally the GPE using the ansatz:
$
\Psi \left( {\bf r},t \right) =
\, c_1(t)\, \psi _1\left( {\bf r} \right) 
+  c_2(t)\, \psi _2\left( {\bf r} \right)    
$,
where $c_{n} = \sqrt{N_{n} (t)} \exp\left( i\theta _{n} (t)\right)$ 
are complex time-dependent amplitudes of the left $n=1$ and 
right $n=2$ condensates (see also \cite{smerzi97}). 
The trial wave functions $\psi_{1,2}\left( {\bf r}\right)$ are orthonormal 
and can be interpreted as approximate ground state solutions of the GPE of 
the left and right wells.
The equations of motion
for the relative population $\eta=(N_2-N_1)/N$ and phase 
$\theta =\theta _2-\theta _1$ between the two symmetric traps
are 
\begin{eqnarray}
\hbar \,\dot{\eta} &=&(2 E_J/N)\sqrt{1-\eta^2}\sin \left( \theta \right) \;,
\label{eq8} \\
\hbar \,\dot{\theta} &=& F\,l_z(t)
-\frac{2 E_J}{N}\frac \eta{\sqrt{1-\eta^2}}
\cos \left( \theta \right) - \frac{N E_c}{2}\eta  \label{eq9} \;,
\end{eqnarray}
where $E_c = 2g\int d{\bf r\,\,}\psi _1({\bf r})^4$ 
is the variational analog of the capacitive energy in SJJ,
while $ E_J =
- N \int d{\bf r} \psi _1({\bf r})
\left[ H_0 +  g N \psi^2_1({\bf r})  \right] \psi _2({\bf r})
$ is the Josephson coupling energy. 
The current-phase relation $I=I_c\sqrt{1-\eta^2}\sin(\theta)$
is directly related to Eq.~(\ref{eq8})
where the critical current is given by $I_c=E_J/\hbar$.
$F\,l_z(t)$ represents the contribution to the 
chemical potential difference in
the two wells due to the laser displacement $l_z$
(after linearizing in $l_z$), and where
$
F =  \int d{\bf r} \left( \psi _1({\bf r})^2 - \psi _2({\bf r})^2 \right)
\frac{\partial}{\partial l_z}V_{laser}
\simeq
  m\omega _0^2\int d{\bf r}\,z\,
\left( \psi _1({\bf r})^2 -
      \psi _2({\bf r})^2   \right) 
$.
The above variational method 
provides a simple and useful interpolating scheme between 
the low interacting limit $N^2 E_c \ll E_J$ and the 
opposite limit $N^2 E_c \gg E_J$.
In the last case, and with $\eta \ll 1$, we recover the 
driven-pendulum phase equation (\ref{eq1}) and the 
critical velocity relations (\ref{eq4}) and (\ref{eq5}) 
with $\hbar \omega_p=\sqrt{E_J\,E_c }$.
In particular, it is legitimate to 
consider the Josephson coupling
as a perturbation, with the 
the phase dynamics entirely determined 
 by the difference in the chemical potentials
$\mu_1(N_1,l_z)$ and $\mu_2(N_2,l_z)$
in the two wells.
In this case $E_c$ corresponds to 
$2 \,\left({\partial \mu_1 }/{\partial N_1} \right)_{l_z}$
and $\hbar^2\omega^2_p=
E_J\left({\partial \mu_1 }/{\partial N_1} \right)_{l_z}$.
The critical velocity is proportional to
the critical current: $v_c = \left(\frac{d\;N_1}{dl_z}\right)^{-1} I_c$,
with
\begin{equation}
\left(\frac{d\;N_1}{dl_z}\right)^{-1} = 
\left( \frac{\partial \mu_1 }{\partial l_z}\right)_{N_1}^{-1}
\left( \frac{\partial \mu_1 }{\partial N_1}\right)_{l_z} 
\label{eq44}
\end{equation}
and  
$\left({\partial \mu_1 }/{\partial l_z} \right)_{N_1}$
being $F/2$ in Eq.(\ref{eq5}).
These derivatives can be computed numerically.
In the Thomas-Fermi (TF) limit they reduce to
\begin{equation}
\left( \frac{\partial \mu_1 }{\partial N_1} \right)_{l_z}
= \frac{g}{V_{TF}}
\end{equation}
and 
\begin{equation}
\left( \frac{\partial \mu_1 }{\partial l_z} \right)_{N_1}
= \frac{1}{V_{TF}}
 \int_{V_{TF}} d{\bf r\,} \frac{\partial}{\partial l_z} V_{laser}
\end{equation}
where $V_{TF}$ is the volume of the region in which $\Psi_1$ 
is different from zero (in the TF approximation).

We make the comparison of  Eqs. (\ref{eq8}) and (\ref{eq9}) 
with a full numerical integration of the 
GPE in an experimentally realistic geometry
relative to the limit $N^2 E_c \gg E_J$. 
In particular, we show that Eq. (\ref{eq5}), derived in 
the limit of  $\eta \ll 1$, still remains a good 
approximation even for $\eta \approx 0.4$.
The details of the numerical calculation are given 
elsewhere \cite{giovanazzi}.

We have considered the JILA setup, with $N=5 \times 10^4$ 
Rb atoms in a cylindrically symmetric harmonic trap, 
having the longitudinal frequency $\omega _0 = 50$ 
s$^{-1}$ and the radial frequency $\omega _r = 17.68$ 
s$^{-1}$.  
The value of the scattering length considered is 
$a=58.19$ $\dot{A}$.
A Gaussian shaped laser sheet  is focused in the center 
of the trap, cutting it into two parts.
We assume that the (longitudinal) $1/e^2$ half-width of 
the laser barrier is $3.5$ $\mu$m and the barrier height 
$V_{0}/\hbar=650$ s$^{-1}$.

Although the lifetime of a trapped condensate can be as 
long as minutes, we have made a quite conservative choice, 
by considering a time scale on the order of one second. 
The possibility to perform experiments on a longer time-scale 
will improve the observability of the phenomena we are 
discussing. 
With this choice of time scale, that corresponds only to few 
plasma oscillations, an adiabatic increase of the velocity 
is not possible, therefore we proceed as follows.
For $t < 0$ the laser is at rest in the middle of the trap, 
$l_z=0$, and the two condensates are in equilibrium.
For $t > 0$ the laser moves across the trap, with constant 
velocity, and the relative atomic population is observed 
at $t_{f}=1~s$. 
With this initial condition, which introduces small plasma 
oscillations in the relative population, it
is expected, in absence of dissipation,
to slightly reduce
the critical current by the numerical factor 
$\approx 0.725$ (see the general properties
of the driven pendulum equation \cite{barone82}). 

In Fig.1 we show the relative condensate population 
$\eta = (N_2 - N_1) / N$, calculated after $1$ second,
for different values of the laser velocity $v$. 
The crosses are the results obtained with the full numerical
integration of the time-dependent GPE (\ref{eq6}).
The dot-dashed line shows the equilibrium values  
$\eta_{eq}$ of the relative population 
calculated with the stationary GPE and with the laser at rest 
in the ''final'' position $l_z = v ~ t_f$. 
The displacement of $\eta(t_{f})$ from $\eta_{eq}$ is a 
measure of the chemical potential difference, being 
$\Delta \mu = \mu_2 - \mu_1\approx N E_c (\eta(t_{f}) - 
\eta_{eq}) / 2$.

For $v <  0.42~\mu m/s$, the atoms tunnel through the 
barrier in order to keep the chemical potential 
difference $\Delta \mu$ locked around zero.
The dc component of the tunneling current is accounted 
for by an averaged constant phase difference between the 
two condensates. 
This is the close analog of the dc Josephson effect in 
superconducting Josephson junctions.   
The small deviations between the dashed line and the crosses are 
due to the presence of plasma oscillations 
(induced by our initial condition).
At $v \approx 0.42~\mu m/s$ there is a sharp transition, 
connected with the crossover from the dc-branch to the 
ac-branch in SJJ. 
For $v >  0.42~\mu m/s$, the phase difference starts running 
and the population difference, after a transient time, 
remains on average fixed. 
A macroscopic chemical potential difference is established 
across the junction. 
In this regime ac oscillations in the population difference 
are observed.
The frequency of such oscillations are approximatively 
given by $\Delta\mu(t)/\hbar$ (not visible in the figure).

%%%%%%%%%%%%%%%%%  F I G U R E %%%%%%%%%%%%%%%%%%%%%%
\begin{figure}[h]
%\vspace{-1.cm}
\begin{center}
\centerline{\psfig{figure=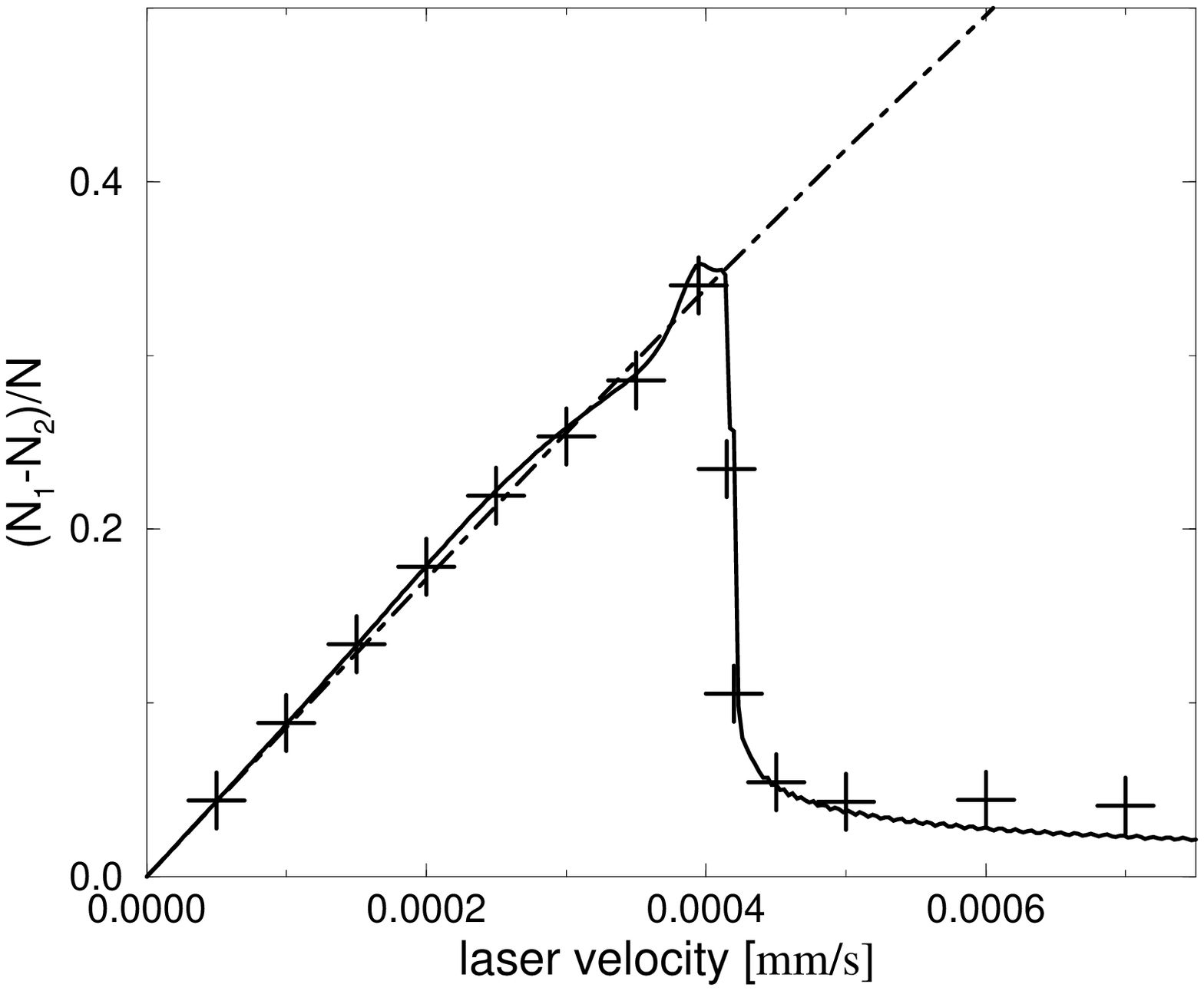,height=6.8cm}}
\end{center}
\vspace{-.5cm}
%%%%%%%%%%%%%%%%%  F I G U R E %%%%%%%%%%%%%%%%%%%%%%
\begin{caption}
{Fractional population imbalance versus the velocity of the
laser creating the weak link. A sharp transition between the
"dc" and the "ac" branches occurs at a barrier critical velocity. 
The solid line and the crosses are the analytical and the numerical
calculations, respectively.
The dashed-dot line represents the static equilibrium value $\eta_{eq}$
calculated with the center of the laser at $v\,t_f$.}
\end{caption}
\end{figure}
%%%%%%%%%%%%%%%%%%%%%%%%%%%%%%%%%%%%%%%%%%%%%%%%%%%%%%

The solid line of Fig.1 corresponds to the solutions of 
Eqs.~(\ref{eq8}) and (\ref{eq9}) in which the value of 
the energy integrals $E_c N/{\hbar} = 2.46~ ms^{-1}$ 
and $E_J/N\hbar = 2.41 \times 10^{-4}~ ms^{-1}$
are chosen in order to give the correct value of $\omega_p
= 2.44 \times 10^{-2}~ ms^{-1}$ and 
$I_c = 12.1~ ms^{-1}$. 
The values $\omega_p$, $I_c$ are calculated numerically 
studying  the frequency of small oscillations around 
equilibrium and the current-phase relation, respectively.
The force integral is $F/\hbar = 1.060~ ms^{-1}~\mu m^{-1}$.
The parameters $\omega_p$, $I_c$ and $F$ are calculated 
with the laser at rest ($v=0$) in $l_z=0$.
Using these values in Eq.~(\ref{eq5}) and taking into 
account the reducing factor $0.725$ we obtain the 
value  $0.407$ $\mu m~ s^{-1}$ for the critical velocity, 
in agreement with the value observed in the simulation.

Small deviations between the variational solutions 
(full line in Fig.1) and the numerical results  
(crosses in Fig.1), above the critical velocity, 
are due to ``level-crossing'' effects.
Numerical results  \cite{giovanazzi} show that when the 
condensate ground state of the ``upper'' well is aligned 
with the excited collective dipole state in the ``lower'' 
well, a finite number of atoms go from the ``upper'' 
well to the ``lower'' well.
Close to this tunneling resonance it is possible to control,
by manipulating the barrier velocity below a fraction of $v_c$, 
the dc flux of atoms
from the ground state condensate in the ``upper'' well
to the longitudinal intrawell collective dipole mode of 
the condensate in the ``lower'' well.
This effect is directly observable in the macroscopic 
longitudinal oscillations of the two condensates
(at frequencies $\approx \omega_0$).

Concerning a possible realization of the phenomenon 
described in this work, we note that for small barrier 
velocities $v$, the motion of the laser sheet with respect 
to the magnetic trap with velocity $v$ or, $viceversa$, 
the motion of the magnetic trap with velocity $-v$, are 
equivalent, there being negligible corrections due to 
different initial accelerations.

Thus far we have discussed the zero temperature limit.
At finite temperature dissipation can arise due to 
incoherent exchange of thermal atoms between the two wells. 
This can be described phenomenologically by including 
a term $- E_c G \dot{\theta} / \omega^2_{p} $ in 
Eq.~(\ref{eq1}) where $G$ is the conductance.
Dissipation will be negligible as long as the 
characteristic time scale $( E_c G )^{-1} \approx 
(20 G/\hbar ) \;s$ is bigger than the time scale of the 
experiment ($\approx 1 s$).

To conclude we note that while it
could be difficult to measure directly the plasma
oscillations, since their amplitude is limited by
$\Delta\eta < \frac{4}{N}\sqrt{\frac{E_J}{E_c}}$,
the macroscopic change in the population 
difference may be easily detected
with standard techniques.
Moreover the framework that we have discussed
 can be easily adapted to investigate
the internal Josephson effect.

Our phenomenological equations 
are similar to the driven pendulum equation 
governing the Josephson effects in SJJs.
As a consequence, within this framework we can study the 
``secondary quantum phenomena'', such as the 
Macroscopic Quantum Tunneling between different local 
minima of the washboard potential
(see for instance \cite{schon90}).

\par
It is a pleasure to thank  L.~{P}.~{P}itaevskii, 
S.~{R}aghavan and S.~{R}.~{S}henoy for many 
fruitful discussions.


\newpage


\begin{thebibliography}{10}

\bibitem[*]{sg} Present Address: Department of Chemical Physics, 
Weizmann Institute of Science, 76100 Rehovot, Israel.

\bibitem{anderson84}
P.~W. Anderson, {\em Basic Notions of Condensed Matter Physics}
  (Benjamin-Cummings, Menlo Park, 1984).

\bibitem{barone82}
A. Barone and G. Paterno, {\em Physics and Applications of the Josephson
  Effect} (Wiley, New York, 1982).

\bibitem{barone99}
{A. Barone, NATO ASI Series {\em Quantum Mesoscopic Phenomena and 
Mesoscopic Devices in Microelectronics}, 
Ankara June 1999, (I.O. Kulik and R. Ellialtioglu Eds.) Kluwer (in press)}.

\bibitem{pereverzev97}
{O. Avenel, and E. Varoquaux, Phys. Rev. Lett. {\bf 55}, 2704 (1985); 
 S.~V. Pereverzev et al., Nature {\bf 388},  449  (1997);
 S. Backhaus, et al., Science {\bf 278},  1435  (1998);
 S. Backhaus, et al., Nature {\bf 392},  687  (1998).}

\bibitem{anderson95}
{M.~H.~Anderson et al., Science {\bf 269}, 198 (1995); 
  K.~B.~Davis, et al., Phys. Rev. Lett. {\bf 75}, 3969 (1995); 
  C.~C.~Bradley, et al., Phys. Rev. Lett. {\bf 75}, 1687 (1995); 
  D. G. Fried, et al., Phys. Rev. Lett. {\bf 81}, 3811 (1998).}

\bibitem{andrews97}
{M.~R. Andrews et al., Science {\bf 275}, 637 (1997).}

\bibitem{hall98}
{D.~S.~Hall et al., {\bf 81}, 1539, 1543 (1998).}

\bibitem{anderson98}
{B.~P. Anderson and M.~A. Kasevich, Science {\bf 282}, 1686 (1998).}

\bibitem{smerzi97}
{A. Smerzi, S. Fantoni, S. Giovanazzi, and S. R. Shenoy, Phys. Rev. Lett.,
  {\bf 79}, 4950 (1997); S. Raghavan, A. Smerzi, S. Fantoni, and S. R. Shenoy,
  Phys. Rev. A, {\bf 59}, 620 (1999).}

\bibitem{milburn97}
{C.~J. Milburn, J. Corney, E.~M. Wright, and D.~F. Walls, 
 Phys. Rev. A {\bf 55},  4318  (1997).}

\bibitem{ruo98}
{J. Ruostekoski and D.J. Walls, Phys. Rev. A {\bf 58} R50 (1998)}
%{J.~I. Cirac, M. Lewenstein, K. M\mbox{\o}lmer, and P. Zoller, 
% Phys. Rev. A {\bf 57},  1208  (1998).}

\bibitem{villain99}
{ P. Villain and M. Lewenstein, Phys. Rev. A {\bf 59}, 2250 (1999).}

\bibitem{zapata98}
{I. Zapata, F. Sols, and A. Leggett, Phys. Rev. A {\bf 57},  R28  (1998).}

\bibitem{williams99} 
{J. Williams, R. Walser, J. Cooper, E. Cornell, and
 M. Holland, Phys. Rev. A {\bf 59},  R31  (1999).}

\bibitem{webb74}
{R.~A.~Webb et al., Phys. Lett {\bf 48}A, 421
 (1974); Phys. Rev. Lett {\bf 33}, 145 (1974);
 A.~J. Leggett, Rev. Mod. Phys. {\bf 47},  331  (1975);
 K. Maki and T. Tsuneto, Prog. Theor. Phys. {\bf 52},  773  (1974).}

\bibitem{pitaevskii61}
{L.~{P}.~{P}itaevskii, {S}ov. {P}hys. {JETP}, {\bf 13}, 451 (1961); 
 E. P. Gross, Nuovo Cimento {\bf 20}, 454 (1961); 
 J. Math. Phys. {\bf 4}, 195 (1963).}

\bibitem{dalfovo98}
{F. Dalfovo, S. Giorgini, L.~{P}.~{P}itaevskii and S. Stringari, 
Rev. Mod. Phys. {\bf 71},  463  (1999).}

\bibitem{giovanazzi}  
{S. Giovanazzi, Ph.D. Thesis, SISSA, Trieste, Italy,
 (1998), unpublished.}

\bibitem{schon90}
{G. Schon, and A. D. Zaikin,
Phys. Rep. {\bf 198},  237  (1999);
P. Silvestrini, B. Ruggiero and A. Esposito, Low Temp. Phys. {\bf 22},
195 (1996).}

\end{thebibliography}
\end{document}